# Effect of carbon vacancies on structural and mechanical properties of stable zirconium carbides: A first principles study

Congwei Xie [a, b, *], Artem R. Oganov [c, d, a, e †], Duan Li [b, a], Tekalign Terfa Debela [a, b], Ning Liu [b, a], Dong Dong [a, b], Qingfeng Zeng [b, a]

[a]*International Center for Materials Discovery, School of Materials Science and Engineering, Northwestern Polytechnical University, Xi'an, Shaanxi 710072, PR China*

[b]*Science and Technology on Thermostructural Composite Materials Laboratory, School of Materials Science and Engineering, Northwestern Polytechnical University, Xi'an, Shaanxi 710072, PR China*

[c]*Skolkovo Institute of Science and Technology,5 Nobel street, Skolkovo 143025, Russia*

[d]*Moscow Institute of Physics and Technology, 9 Institutskiy Lane, Dolgoprudny City, Moscow Region 141700, Russia*

[e]*Department of Geosciences and Center for Materials by Design, Stony Brook University, Stony Brook, New York 11794, USA*

* xiecw1021@mail.nwpu.edu.cn

† artem.oganov@stonybrook.edu

## Abstract

By using evolutionary algorithm USPEX, we have predicted a number of stable zirconium carbides. In addition to the well-known rocksalt-type stoichiometric ZrC ($Fm\bar{3}m$), present prediction also identifies five stable substoichiometric zirconium carbides adopting rocksalt-type structures with ordered carbon vacancies, $Zr_8C_7$ ($P\bar{1}$), $Zr_6C_5$ ($C2/m$), $Zr_5C_4$ ($P\bar{1}$), $Zr_3C_2$ ($C2/m$), and $Zr_2C$ ($Fd\bar{3}m$). The effects of carbon vacancies on structural and mechanical properties are investigated. We highlight that the distribution of carbon vacancies has significant influence on volume, Pugh's ratio, and hardness. We further propose that hardness can be enhanced by replacing carbon vacancies with suitable elements, in particular nitrogen and oxygen.

Transition metal carbides (TMCs) have a broad range of technological applications due to their outstanding hardness, extremely high melting points, good wear and corrosion resistance, and excellent thermal conductivity[1-6]. Structures of most TMCs are based on close packing of metal atoms, with C atoms occupying octahedral voids. In substoichiometric TMCs, not all octahedral voids are occupied by C atoms. Because of carbon vacancies, properties of substoichiometric TMCs can be very different from their stoichiometric counterparts[5-9].

In order to understand the properties of substoichiometric TMCs more deeply, we need to clarify the crystal structures of them. Experimentally, crystal structures of substoichiometric TMCs cannot be easily characterized because of similarities to their stoichiometric forms, except the concentration and distribution of carbon vacancies. Due to the weak atomic scattering factor of carbon, the X-ray diffraction patterns between substoichiometric TMCs and their stoichiometric forms are almost undistinguishable[5,8]. Theoretical methods, in particular the order-parameter functional method[10-12], and Monte Carlo calculations[13], in many cases are reliable (e.g., investigations for substoichiometric TMCs $Ti_2C$, $Zr_2C$, and $V_6C_5$ etc.[5,6,8,14]), but in some cases are questionable (e.g., see examples in Refs. 10-13). The state-of-the-art evolutionary algorithm (EA) for structure prediction, performing global search and implemented in the USPEX code is a promising approach to solve this problem[15,16]. Using USPEX, many stable substoichiometric TMCs have been found, e.g. in systems Ta-C[17], Ti-C[18], and Hf-C[19].

Zirconium carbide is a technologically useful TMC material for various applications. According to previous studies, the concentration of carbon vacancies in zirconium carbide can be up to 50% [8-12]. Carbon vacancies are generally disordered in substoichiometric zirconium carbides at high temperatures. After a long-time annealing, these substoichiometric zirconium carbides prefer to adopt ordered carbon vacancies forms[8,9]. Gusev et al. did some experiments to verify the existence of substoichiometric zirconium carbides with ordered carbon vacancies[11,12]. According to their experimental studies, $Zr_6C_5$, $Zr_3C_2$, and $Zr_2C$ are stable, and they also proposed some ordered structures for these three substoichiometric zirconium carbides based on the order-parameter functional method[10-12]. However, there remain two unsolved issues: (1) all stable substoichiometric zirconium carbides are not explored; and (2) structures proposed by order-parameter functional method are not always stable. Therefore, a global search for all the stable substoichiometric zirconium carbides is desired.

In the present study, we used the EA method to predict stable zirconium carbides at zero pressure and zero temperature. We report five stable substoichiometric zirconium carbides with ordered carbon vacancies, $Zr_8C_7$ ($P\bar{1}$), $Zr_6C_5$ ($C2/m$), $Zr_5C_4$ ($P\bar{1}$), $Zr_3C_2$ ($C2/m$), and $Zr_2C$ ($Fd\bar{3}m$). Based on the predicted structures, we further investigated the effects of carbon vacancies on the structural and mechanical properties of substoichiometric zirconium carbides. Our study shows that the relationship between the concentration of carbon vacancies with bulk and shear moduli is almost linear, whereas that of hardness is nonlinear and monotonic. Interestingly, volume and Pugh's ratio have a maximum at carbon vacancies concentration around 1/6. We suggest that these non-trivial trends are associated with the evolution of carbon vacancies distribution. Our present study may inspire further experiments on tailoring properties of zirconium carbides by altering the concentration and distribution of carbon vacancies.

Let us firstly focus on searching for the stable zirconium carbides. We performed EA prediction using USPEX code[15,16]. Stability of a compound is determined using the thermodynamic convex hull construction. For every candidate structure generated by USPEX, total energy calculations and structure relaxations were carried out using density functional theory (DFT[20]) within the Perdew-Burke-Ernzerhof generalized gradient approximation (GGA[21]), as implemented in the VASP code[22]. The all-electron projector-augmented wave method[23], with a plane-wave kinetic energy cutoff of 560 eV, and uniform $k$-point meshes with reciprocal-space resolution of $2\pi \times 0.06\,\mathrm{\AA}^{-1}$, were used. These settings enable excellent convergence of the energy differences, stress tensors, and structural parameters. Denser $k$-point meshes in reciprocal space with a resolution of $2\pi\times 0.03\,\text{Å}^{-1}$ were used for detailed calculations of properties.

Results of the global EA search for stable zirconium carbides are shown in Fig. 1. In addition to the most common rocksalt stoichiometric ZrC ($Fm\bar{3}m$) and the previously confirmed substoichiometric $Zr_2C$ ($Fd\bar{3}m$) structures[8,10,11], our calculations have identified four other stable ordered substoichiometric structures: $Zr_8C_7$ ($P\bar{1}$), $Zr_6C_5$ ($C2/m$), $Zr_5C_4$ ($P\bar{1}$), and $Zr_3C_2$ ($C2/m$). Additionally, the $R\bar{3}m$ phase of $Zr_8C_5$ possessing the same topology as the stable $Ti_8C_5$[24] is also discovered. However, the $R\bar{3}m$ phase is metastable as its enthalpy of formation lies above the convex hull. Compared with previous results calculated using the order-parameter functional method by Gusev *et al.*[10,11] (see Fig. 1), the present EA search has indentified the same structures for $Fd\bar{3}m$ $Zr_2C$ and $C2/m$ $Zr_6C_5$, but has discovered lower-energy structures for $Zr_8C_7$ ($P\bar{1}$), $Zr_3C_2$ ($C2/m$), and $Zr_4C_3$ ($I\bar{4}3m$). Besides, a stable substoichiometric zirconium carbide $Zr_5C_4$ ($P\bar{1}$) – a composition never reported before – has been found. We have found that this $P\bar{1}$ structure is also adopted by thermodynamically stable $Ti_5C_4$ and $Hf_5C_4$ (see Figs. S1&S2 in the Supplementary Material). Detailed structural information for these predicted stable zirconium carbides is listed in Table SI in the Supplementary Material.

To ensure the dynamical stability of the predicted compounds, a finite displacement method, as implemented in the PHONOPY code[25], was used to obtain phonon spectra. For a given crystal, each atom, in turn, is displaced by a small amount, and the forces induced on all the atoms in the crystal are calculated by VASP and used to construct the force constant matrix. The VASP calculation is performed on a supercell with a sufficiently large number of atoms. From the force constants matrix we compute the dynamical matrix at any chosen $q$-point of Brillouin zone, and the diagonalisation of the dynamical matrix provides squares of the phonon frequencies[25]. As shown in Fig. S3 in the Supplementary Material, no imaginary phonon frequencies have been found throughout the Brillouin zone, suggesting dynamical stability of these substoichiometric phases.

The elastic constants, directly related to the mechanical stability of a material, were also calculated by VASP. We have also found these substoichiometric zirconium carbides to be mechanically stable because their elastic constants satisfy the mechanical stability criteria[26] (see Table SII in the Supplementary Material).

In all of these stable zirconium carbides, C atoms occupy octahedral voids in the cubic close packing of Zr atoms, and all the C atoms are six-coordinate. In the stoichiometric $Fm\bar{3}m$ ZrC, Zr atoms are six-coordinate, while in substoichiometric zirconium carbides Zr coordination number varies with the number of missing carbon atoms. For stable substoichiometric zirconium carbides $Zr_8C_7$ ($P\bar{1}$), $Zr_6C_5$ ($C2/m$), $Zr_5C_4$ ($P\bar{1}$), $Zr_3C_2$ ($C2/m$), and $Zr_2C$ ($Fd\bar{3}m$), one-of-eight, -six, -five, -three, and -two octahedral interstitial positions are unoccupied, respectively. As a result, their Zr coordination numbers gradually decrease (see Table SI in the Supplementary Material).

Zr coordination number depicts the information of the carbon vacancy distribution: if the coordination number of Zr atom is not less than the threshold value five, the carbon vacancies are nonadjacent. If the coordination number of Zr atom is smaller than this threshold value, carbon vacancies near Zr atom will be grouped. Fig. 2(a)-(f) shows different arrangements of carbon vacancies as visualized by the VESTA software[27]. Carbon vacancies in both $P\bar{1}$ $Zr_8C_7$ and $C2/m$ $Zr_6C_5$ are nonadjacent; in fact, $Zr_6C_5$ is the zirconium carbide with the maximum carbon vacancy concentration that adopts a nonadjacent carbon vacancy structure when all the Zr atoms are five-coordinate. As carbon vacancy concentration increases, the coordination number of Zr becomes smaller than five, and thus carbon vacancies get grouped. As shown in Fig. 2(d)-(f), the stable $P\bar{1}$ $Zr_5C_4$ phase possesses paired adjacent carbon vacancies, and carbon vacancies are interconnected in $C2/m$ $Zr_3C_2$ and $Fd\bar{3}m$ $Zr_2C$.

The relationships between average volume per Zr atom and the concentration of carbon vacancies are shown in Fig. 2(g). As a whole, the occurrence of carbon vacancies doesn't change the volume per Zr atom a lot (within 0.5% for all the stable substoichiometric zirconium carbides). Interestingly, as the concentration of carbon vacancies increases, the volume per Zr atom initially expands and then shrinks. We associate this unusual trend with the evolution of carbon vacancies; that is nonadjacent carbon vacancies contribute to volume expansion, while adjacent carbon vacancies lead to volume shrinkage. This can be interpreted by the redistribution of bond lengths due to carbon vacancies (see Fig. S4 in the Supplementary Material). The lattice relaxation around carbon vacancies results in a strongly anisotropic distortion of the structure (see the widely scattered Zr-C and Zr-Zr bonds length in the case of substoichiometric zirconium carbides). In comparison to the bonds length in $Fm\bar{3}m$ ZrC, some Zr-C and Zr-Zr bonds in $P\bar{1}$ $Zr_8C_7$ and *C2/m* $Zr_6C_5$ are shortened (S-bond), while the rest are lengthened (L-bond). With increasing carbon vacancies concentration, more and more carbon vacancies will be grouped, and the lattice relaxation around grouped vacancies will result in shortening of most of bonds (both S-bond and L-bond). Due to this, there is a trend that the number of S-bonds increases while that of L-bonds decreases. These could explain why nonadjacent carbon vacancies can contribute to volume expansion, while the occurrence of adjacent carbon vacancies leads to volume shrinkage.

The effect of carbon vacancies on mechanical properties (bulk modulus, shear modulus, Pugh's ratio, and Vickers hardness) has also been investigated in this work. The mechanical properties (bulk modulus $B$ and shear modulus $G$) of a polycrystalline material, which can be considered to be homogeneous, were evaluated using the Hill averaging scheme[28-29]. With $B$ and $G$, the Vickers hardness $H_v$ of a material can be obtained according to Chen's empirical model[30]:

$$H_v = 2(k^2G)^{0.585} - 3, \quad (1)$$

where $k$ (equals to $G/B$) is Pugh's ratio[31], and all moduli are in GPa.

The relationships between these mechanical properties and the concentration of carbon vacancies are shown in Fig. 3 (denoted by solid squares and black curves). The bulk modulus and shear modulus vary linearly with the concentration of carbon vacancies, indicating that at a given concentration the distribution of carbon vacancies has a little effect on these moduli. However, the Pugh's ratio first increases and then decreases with increasing carbon vacancies concentration. Unlike for the bulk and shear moduli, the dependence of hardness on the carbon vacancies concentration is nonlinear and monotonic. Just as for the volume, we suppose the distribution of carbon vacancies can affect the Pugh's ratio and hardness of materials. The nonadjacent carbon vacancies enhance Pugh's ratio while the adjacent carbon vacancies decrease Pugh's ratio. The nonadjacent carbon vacancies also have a smaller contribution to the decrease of hardness than adjacent carbon vacancies. To check this hypothesis, $Zr_{10}C_9$ and $Zr_6C_5$ structures (denoted by the open squares in Fig. 3) with adjacent carbon vacancies were constructed basing on $P\bar{1}$ $Zr_5C_4$ and $C2/m$ $Zr_3C_2$ structures, respectively. In contrast to $P\bar{1}$ $Zr_8C_7$ with nonadjacent carbon vacancies, $P\bar{1}$ $Zr_{10}C_9$ has a lower carbon vacancies concentration, but lower Pugh's ratio. $P\bar{1}$ $Zr_{10}C_9$ is also softer than $P\bar{1}$ $Zr_8C_7$ and slightly harder than $P\bar{1}$ $Zr_5C_4$ with adjacent carbon vacancies. Most obviously, $C2/m$ $Zr_6C_5$ with adjacent carbon vacancies has lower Pugh's ratio and hardness than $C2/m$ $Zr_6C_5$ with nonadjacent vacancies. Thus, we conclude that such unusual changes in Pugh's ratio and hardness are related to the evolution of the carbon vacancies distribution.

The mechanical properties of carbides are directly related to their electronic structures. To relate the mechanical properties with chemical bonding, we have computed the electronic density of states (DOS) and the crystal orbital Hamilton populations (COHP) using VASP and LOBSTER programs[32-33], respectively.

The electronic density of states (DOS) and the atom resolved partial density of states (PDOS) of all stable zirconium carbides are shown in Fig. 4 (a). All the stable zirconium carbides are metallic due to their finite electronic DOS at the Fermi level. These finite electronic DOS at the Fermi level are mainly contributed by zirconium's d electron. For all the stable zirconium carbides, zirconium and carbon atoms form strong covalent bonds due to a clear overlap of zirconium's d electron and carbon's p electron curves. The crystal orbital Hamilton populations (COHP) which are helpful to identify the bonding and anti-bonding interactions were computed and shown in Fig. 4 (b). In $Fm\bar{3}m$ ZrC, Zr-C bonding states are located below the Fermi level, and their antibonding counterparts appear just above the Fermi level. When carbon vacancies are created in substoichiometric zirconium carbides, Zr-C antibonding states appear below the Fermi level and move to the low energy range with the carbon vacancies concentration increasing.

The occupation of these Zr-C antibonding levels may explain the mechanical properties of these substoichiometric zirconium carbides. Our hypothesis for explaining their bulk and shear moduli is that with increasing concentration of carbon vacancies, Zr-C bonding weakens. This hypothesis is supported by the integrated crystal orbital Hamilton population (ICOHP) values of bonding Zr-C in all of the stable zirconium carbides. The ICOHP value of a bond can be as an indicator of the bond strength. Because of the appearance of Zr-C antibonding levels below the Fermi level, the average ICOHP value of Zr-C interactions observed in $P\bar{1}$ $Zr_8C_7$ (-2.37 eV/bond), $C2/m$ $Zr_6C_5$ (-2.41 eV/bond), $P\bar{1}$ $Zr_5C_4$ (-2.37 eV/bond), $C2/m$ $Zr_3C_2$ (-2.40 eV/bond), and $Fd\bar{3}m$ $Zr_2C$ (-2.24 eV/bond), are smaller than that of in $Fm\bar{3}m$ ZrC (-4.70 eV/bond). Due to this decline in the strength of Zr-C bonds, the bulk modulus and shear modulus in the stable substoichiometric zirconium carbides decrease.

The anomalous Pugh's ratio and hardness can also be explained by ICOHP values. There is an anti-correlation between ICOHP values of metal-metal bonds and Pugh's ratio in pyrite-type transition-metal pernitrides[34]. We found that such correlation between a specific bond with mechanical properties could be applied here. The Zr-Zr ICOHP values of $P\bar{1}$ $Zr_8C_7$ (-0.58 eV/bond), $C2/m$ $Zr_6C_5$ (-0.59 eV/bond), and $P\bar{1}$ $Zr_5C_4$ (-0.58 eV/bond) are lower than that of $Fm\bar{3}m$ ZrC (-0.82 eV/bond).

Conversely, ICOHP values of Zr-Zr bond in *C2/m* $Zr_3C_2$ (-1.30 eV/bond) and $Fd\bar{3}m$ $Zr_2C$ (-1.31 eV/bond) are higher than in $Fm\bar{3}m$ ZrC. Thus, Pugh's ratio of $Zr_8C_7$ ($P\bar{1}$), $Zr_6C_5$ (*C*2/*m*), and $Zr_5C_4$ ($P\bar{1}$) are larger than that of $Fm\bar{3}m$ ZrC, while for $Zr_3C_2$ (*C*2/*m*) and $Zr_2C$ ($Fd\bar{3}m$) are lower. Thus, Zr-Zr ICOHP values explain the anomalous changes observed in Pugh's ratio and hardness of zirconium carbides.

Once again, we proved that carbon vacancies have a significant effect on the mechanical properties of zirconium carbides. If carbon vacancy sites are occupied by other elements (such as N, B, O, etc), which is thermodynamically favorable at high temperatures, an improvement of the mechanical properties of zirconium carbides may be realized. An enhanced hardness has been reported in transition-metal carbonitrides[35-37]. That means nitriding substoichiometric zirconium carbides will effectively improve hardness. Here, we try to study the impact of B, N and O on mechanical properties. In principle, we need to perform a global search for these ternary systems, which would require calculations for many compositions and, for each composition, many configurations. However, this is computationally prohibitive and our purpose is not to study all possible compositions and their lowest-energy configurations. We aim at providing some information for further study by studying some possible compositions and configurations.

Based on the predicted stable substoichiometric zirconium carbides $Zr_8C_7$ ($P\bar{1}$), $Zr_6C_5$ (*C2/m*), and $Zr_5C_4$ ($P\bar{1}$), we constructed three $ZrC_{1-x}T_x$ ($x$=1/8, 1/6, 1/5) structures for each type of element T (T= B, N, O) and calculated their mechanical properties. These structures are reasonable in terms of their negative formation enthalpies (see Table SIII in the Supplementary Material). The relationships between mechanical properties and T content in such compounds $ZrC_{1-x}T_x$ are shown in Fig.3 (B, N, and O doping are denoted by green, blue, and red curves, respectively). We can see that only the bulk modulus is improved when doping with B, while enhanced bulk modulus, shear modulus, and hardness are found for doping with N and O. Interestingly, hardness has a maximum in the case of N and O. That means the hardness of some ternary systems can be higher than those of their corresponding binary systems.

In summary, evolutionary structure prediction using the USPEX code has been used to solve the controversy regarding the structures and stoichometries of stable zirconium carbides. In addition to two previously known phases, $Zr_2C$ ($Fd\bar{3}m$) and $Zr_6C_5$ ($C2/m$), here we have found three hitherto unknown substoichiometric phases, $Zr_8C_7$ ($P\bar{1}$), $Zr_5C_4$ ($P\bar{1}$), and $Zr_3C_2$ ($C2/m$). All of the stable substoichiometric structures have ordered carbon vacancies. When the concentration of carbon vacancy is lower than 1/6, substoichiometric zirconium carbides prefer to adopt structures with nonadjacent vacancy distribution. As carbon vacancy concentration increases, the initially nonadjacent carbon vacancies gradually become closer and group together. Such evolution of the distribution of carbon vacancies affects structural and mechanical properties, especially volume, Pugh's ratio, and hardness. Further, we studied the mechanical properties of $ZrC_{1-x}T_x$ ($x$ = 1/8, 1/6, and 1/5, and T = B, N, and O) constructed by replacing carbon vacancies in $P\bar{1}$ $Zr_8C_7$, $C2/m$ $Zr_6C_5$, and $P\bar{1}$ $Zr_5C_4$ with T. An enhanced hardness can be found in the case of N and O. This indicates that nitriding (or oxygen permeating) substoichiometric zirconium (or hafnium) carbides may improve their properties, such as hardness.

**Acknowledgements**

C. W. Xie thanks Dr. H. Y. Niu for discussing. This work was supported by the Natural Science Foundation of China (Nos. 51372203 and 51332004), the Foreign Talents Introduction and Academic Exchange Program of China (No. B08040), DARPA (Grant No. W31P4Q1210008), and the Government of Russian Federation (Grant No. 14.A12.31.0003). The authors also acknowledge the High Performance Computing Center of NWPU for the allocation of computing time on their machines.

**Figure captions**

**Fig. 1:** Thermodynamic convex hull of the Zr-C system. Circles denote the results of this study; squares indicate calculations by Gusev et al.

**Fig. 2:** (a)-(f) Atomic arrangement of stable zirconium carbides. Green spheres denote Zr atoms; brown spheres, C atoms; squares, carbon vacancies. (g) Relationships between the carbon vacancy concentration and volume per Zr atom (in $\mathring{A}^3$) for all stable zirconium carbides.

**Fig. 3:** Mechanical properties (bulk modulus, shear modulus, Pugh's ratio and Vickers hardness) of $ZrC_{1-x}$ ($x$ is the concentration of carbon vacancies; $x$=0, 1/10, 1/8, 1/6, 1/5, 1/3, and 1/2) and $ZrC_{1-x}T_x$ (T=B, N, and O; $x$=0, 1/8, 1/6, and 1/5) compounds. Solid squares denote stable $ZrC_{1-x}$ compounds; open squares, constructed $ZrC_{1-x}$ compounds; black curves, the fitted curves for stable $ZrC_{1-x}$ compounds; green curves, the fitted curves for constructed $ZrC_{1-x}B_x$ compounds; blue curves, the fitted curves for constructed $ZrC_{1-x}N_x$ compounds; red curves, the fitted curves for constructed $ZrC_{1-x}O_x$ compounds;

**Fig. 4:** The calculated (a) Density of states (DOS) (b) Crystal orbital Hamiltonian populations (-COHP) curves of Zr-C and Zr-Zr interactions for all the stable zirconium carbides.

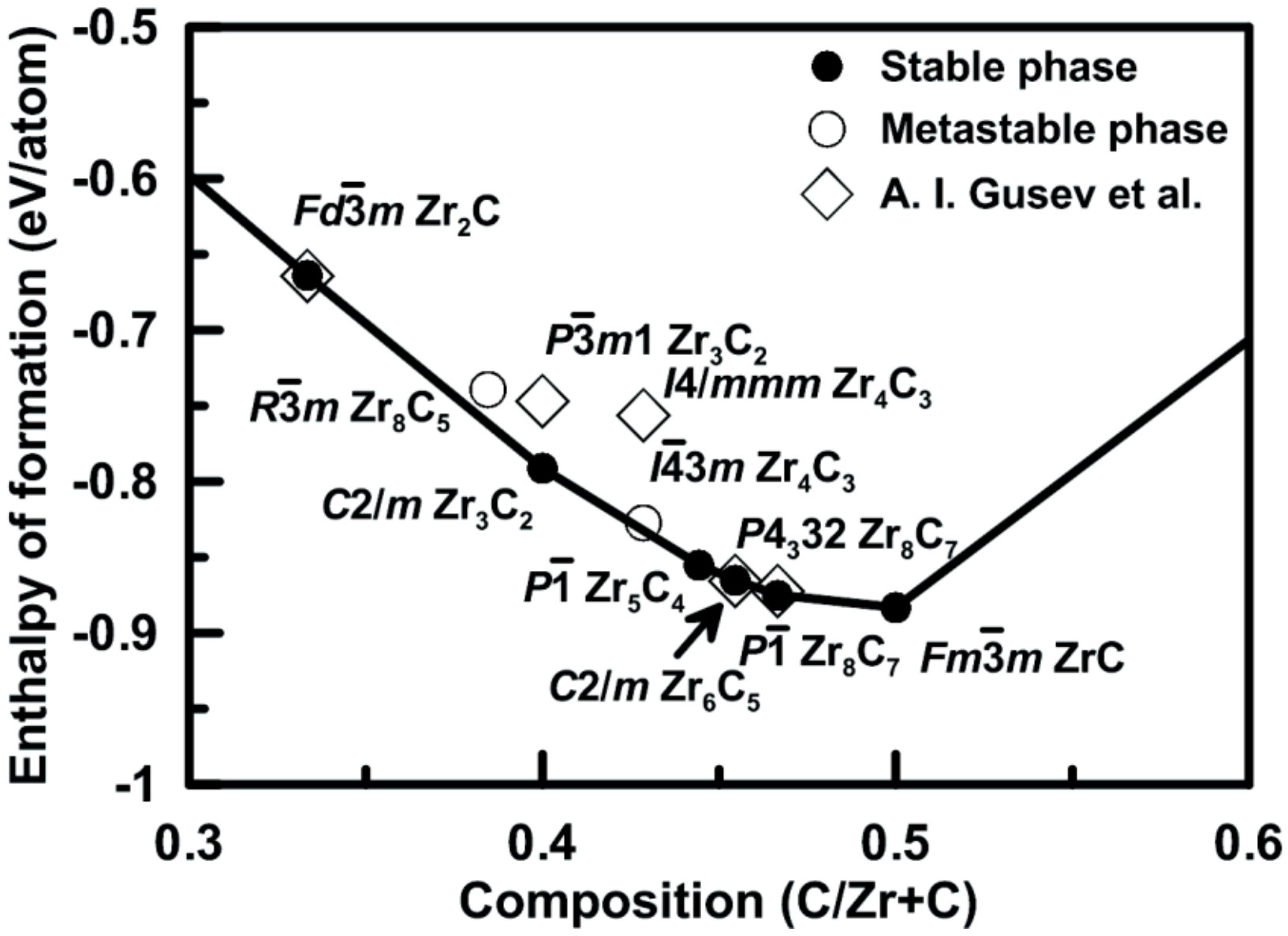

**Fig. 1:** Thermodynamic convex hull of the Zr-C system. Circles denote the results of this study; squares indicate calculations by Gusev et al.

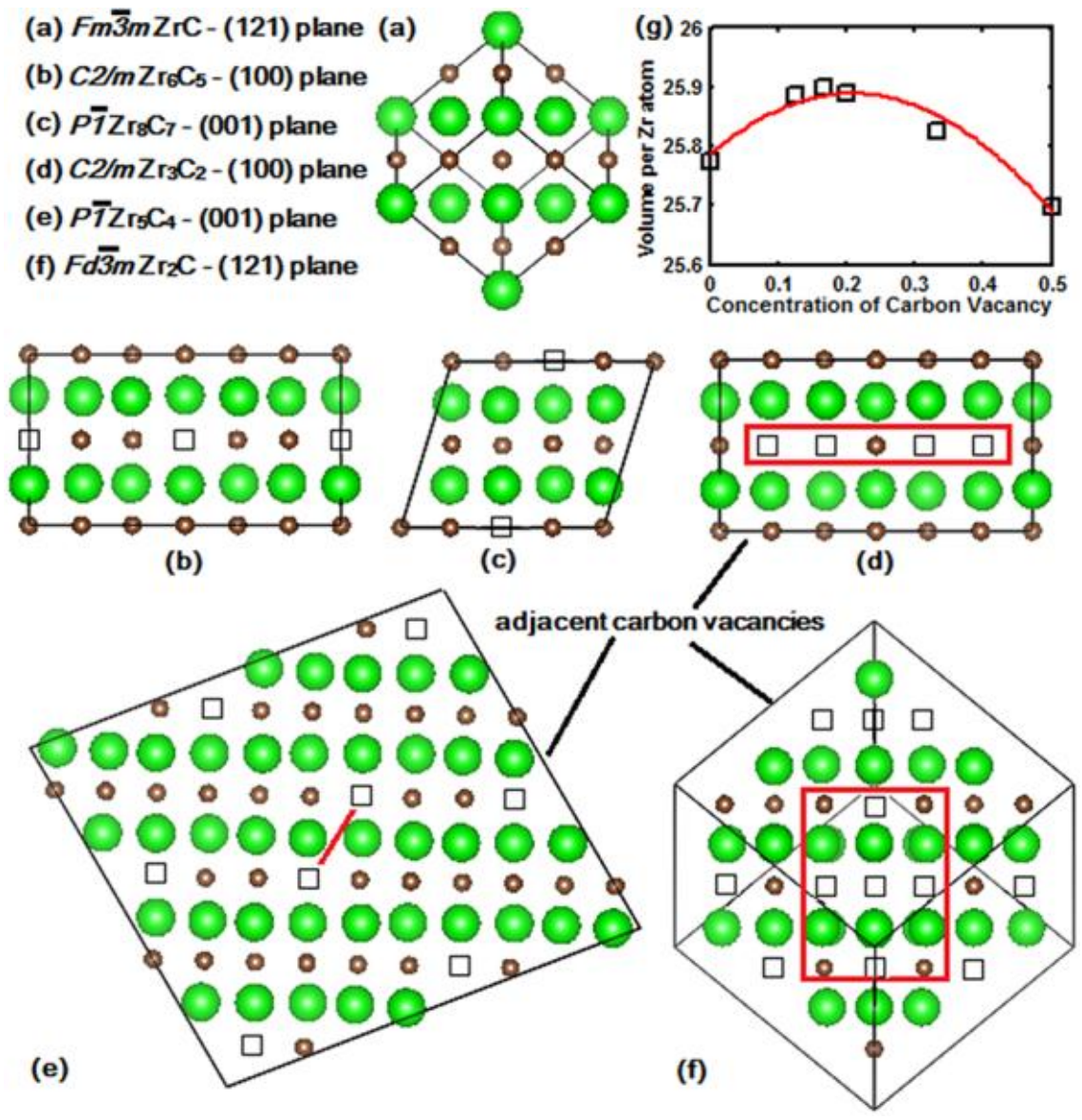

**Fig. 2:** (a)-(f) Atomic arrangement of stable zirconium carbides. Green spheres denote Zr atoms; brown spheres, C atoms; squares, carbon vacancies. (g) Relationships between the carbon vacancy concentration and volume per Zr atom (in $\mathring{A}^3$) for all stable zirconium carbides.

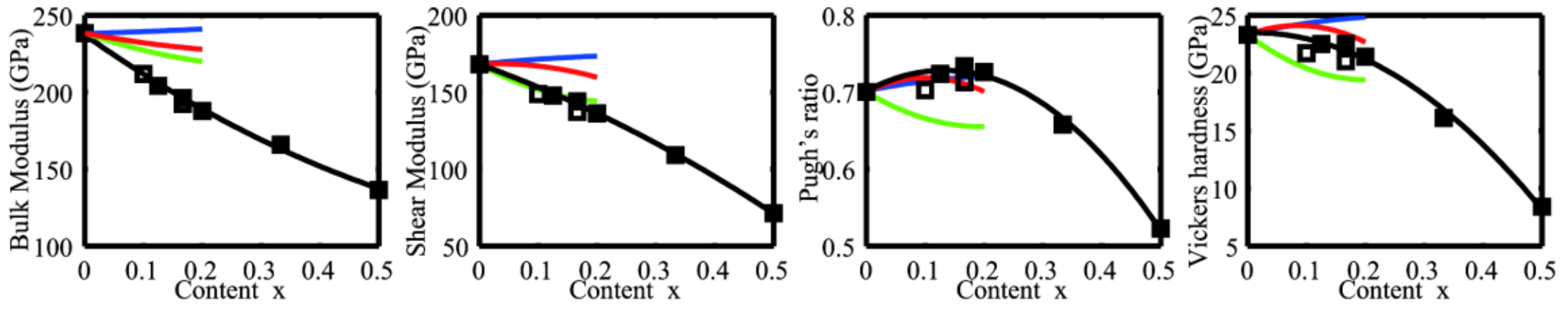

**Fig. 3:** Mechanical properties (bulk modulus, shear modulus, Pugh's ratio and Vickers hardness) of $ZrC_{1-x}$ ($x$ is the concentration of carbon vacancies; $x$=0, 1/10, 1/8, 1/6, 1/5, 1/3, and 1/2) and $ZrC_{1-x}T_x$ (T=B, N, and O; $x$=0, 1/8, 1/6, and 1/5) compounds. Solid squares denote stable $ZrC_{1-x}$ compounds; open squares, constructed $ZrC_{1-x}$ compounds; black curves, the fitted curves for stable $ZrC_{1-x}$ compounds; green curves, the fitted curves for constructed $ZrC_{1-x}B_x$ compounds; blue curves, the fitted curves for constructed $ZrC_{1-x}N_x$ compounds; red curves, the fitted curves for constructed $ZrC_{1-x}O_x$ compounds.

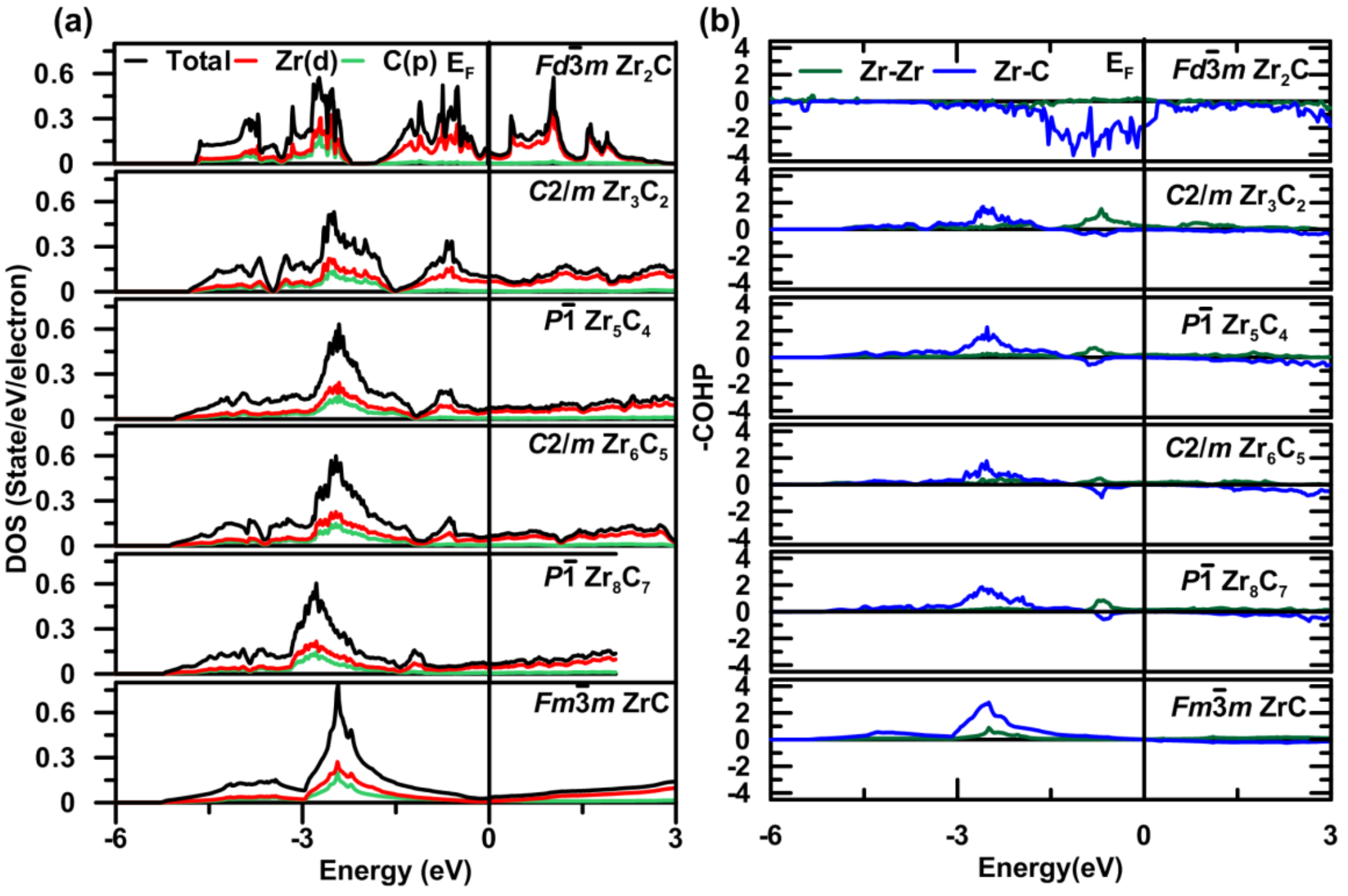

**Fig. 4:** The calculated (a) Density of states (DOS) (b) Crystal orbital Hamiltonian populations (-COHP) curves of Zr-C and Zr-Zr interactions for all the stable zirconium carbides.

## Supplementary Material

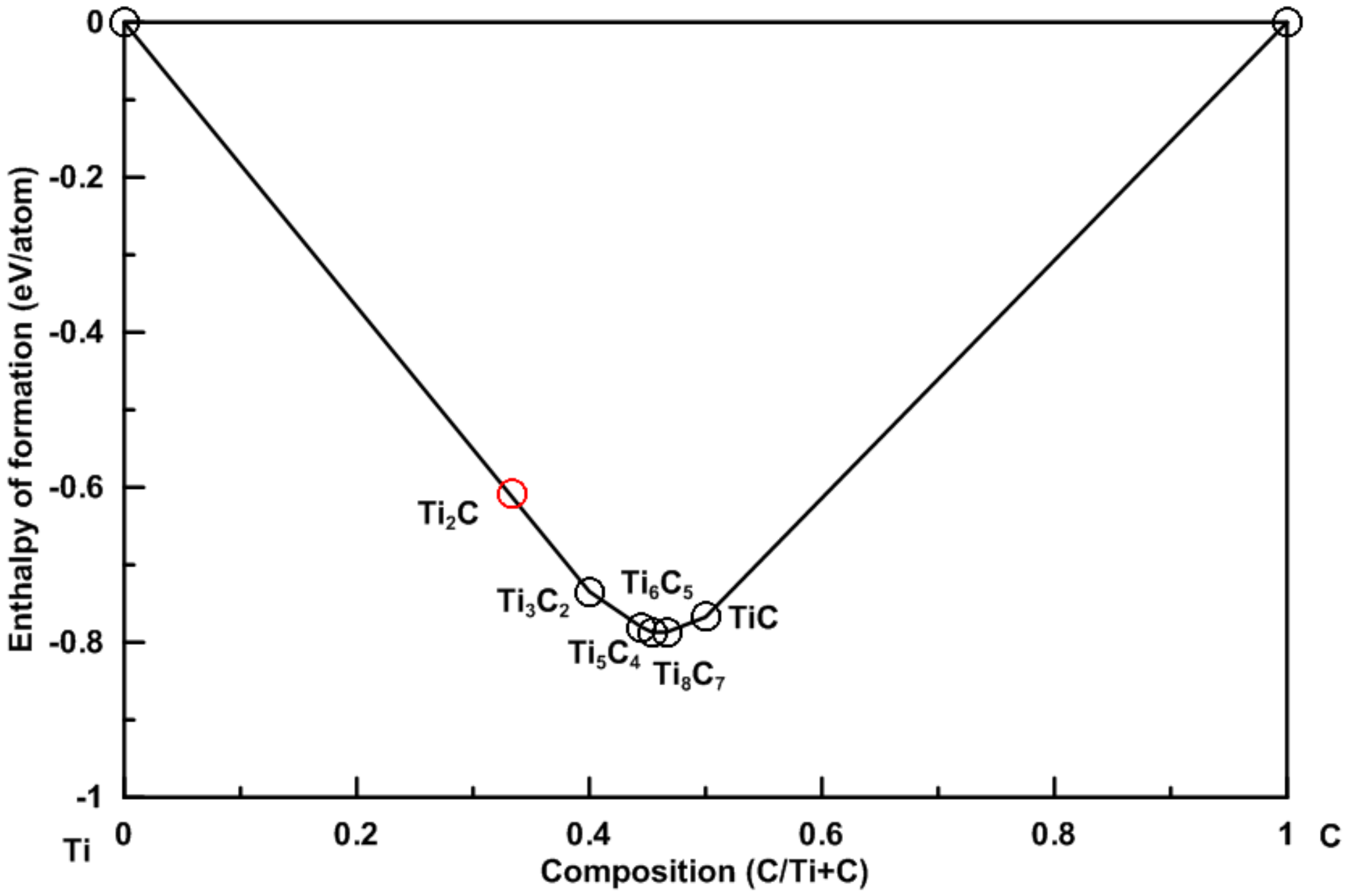

Fig. S1: The thermodynamic convex hull of Ti-C system. Black circles denote the stable phases, while red circle the metastable phase.

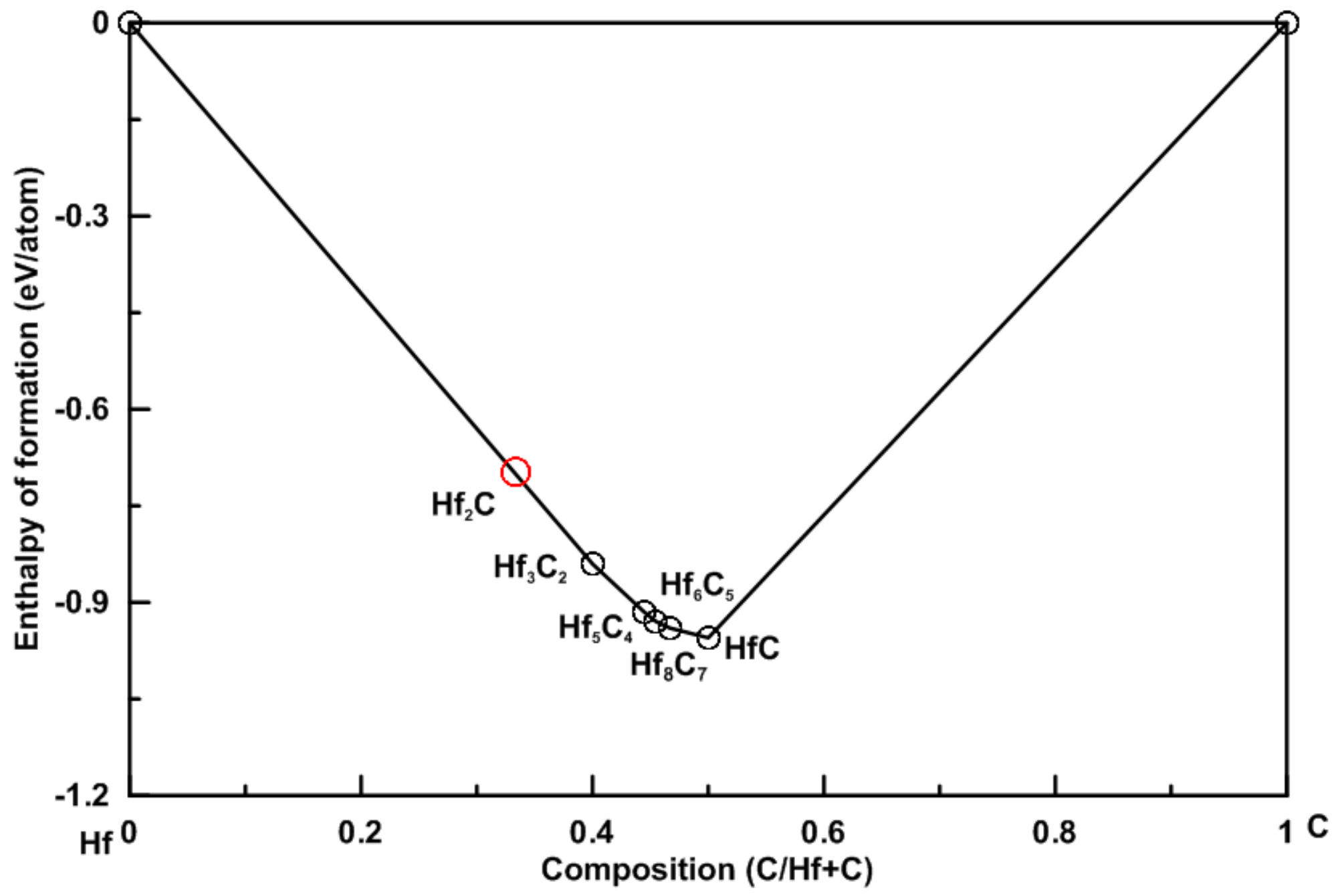

Fig. S2: The thermodynamic convex hull of Hf-C system. Black circles denote the stable phases, while red circle the metastable phase.

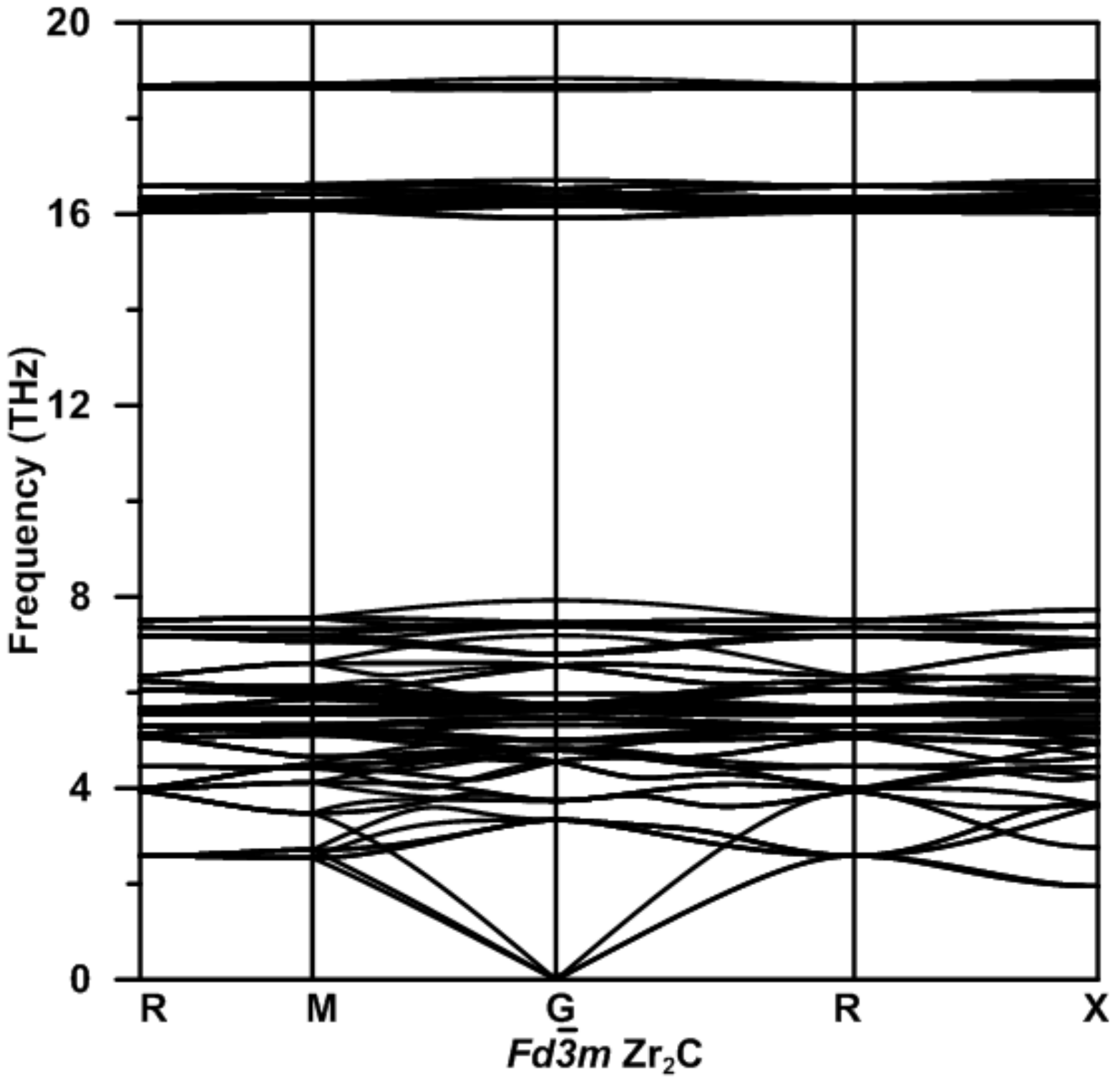

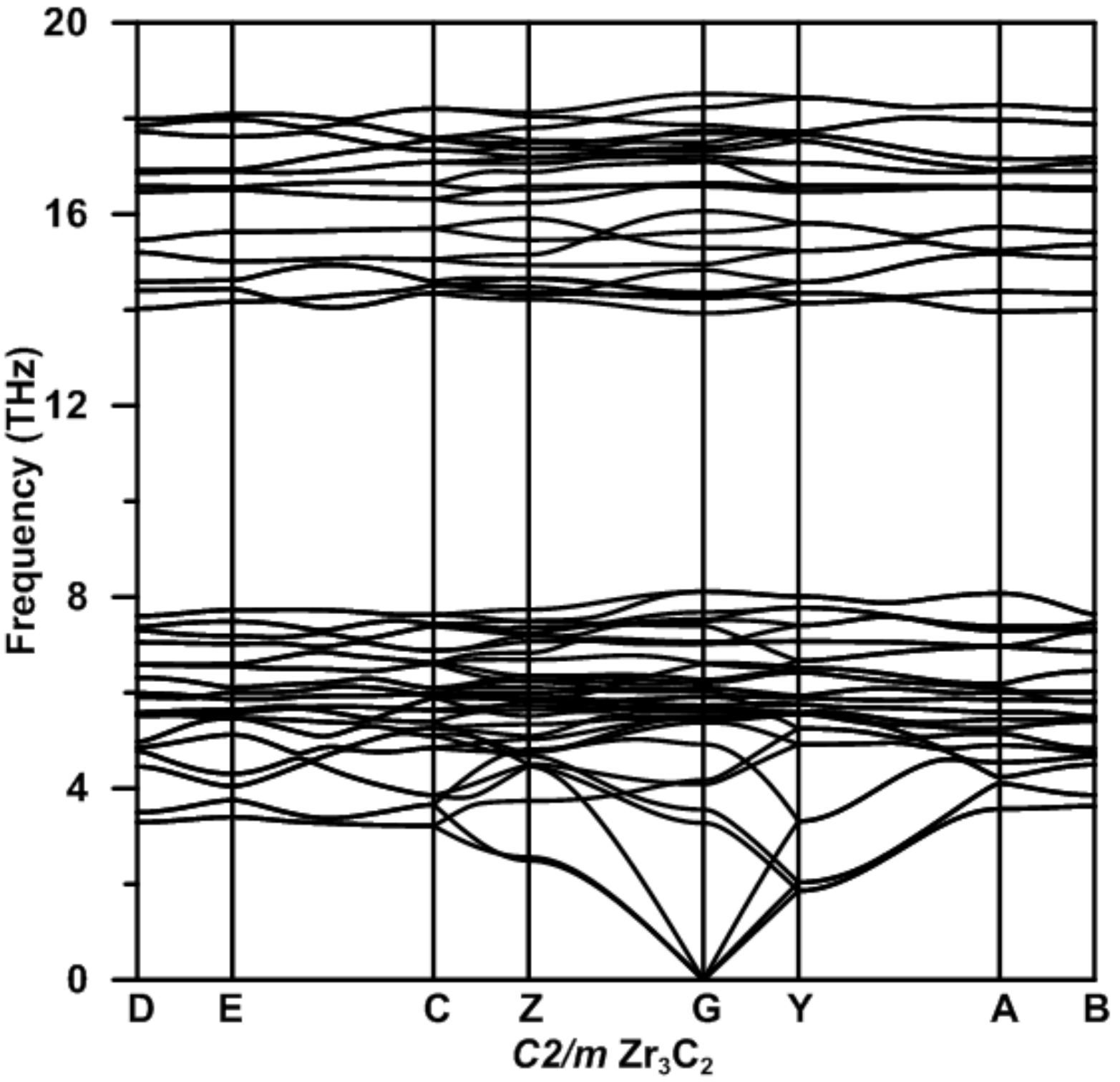

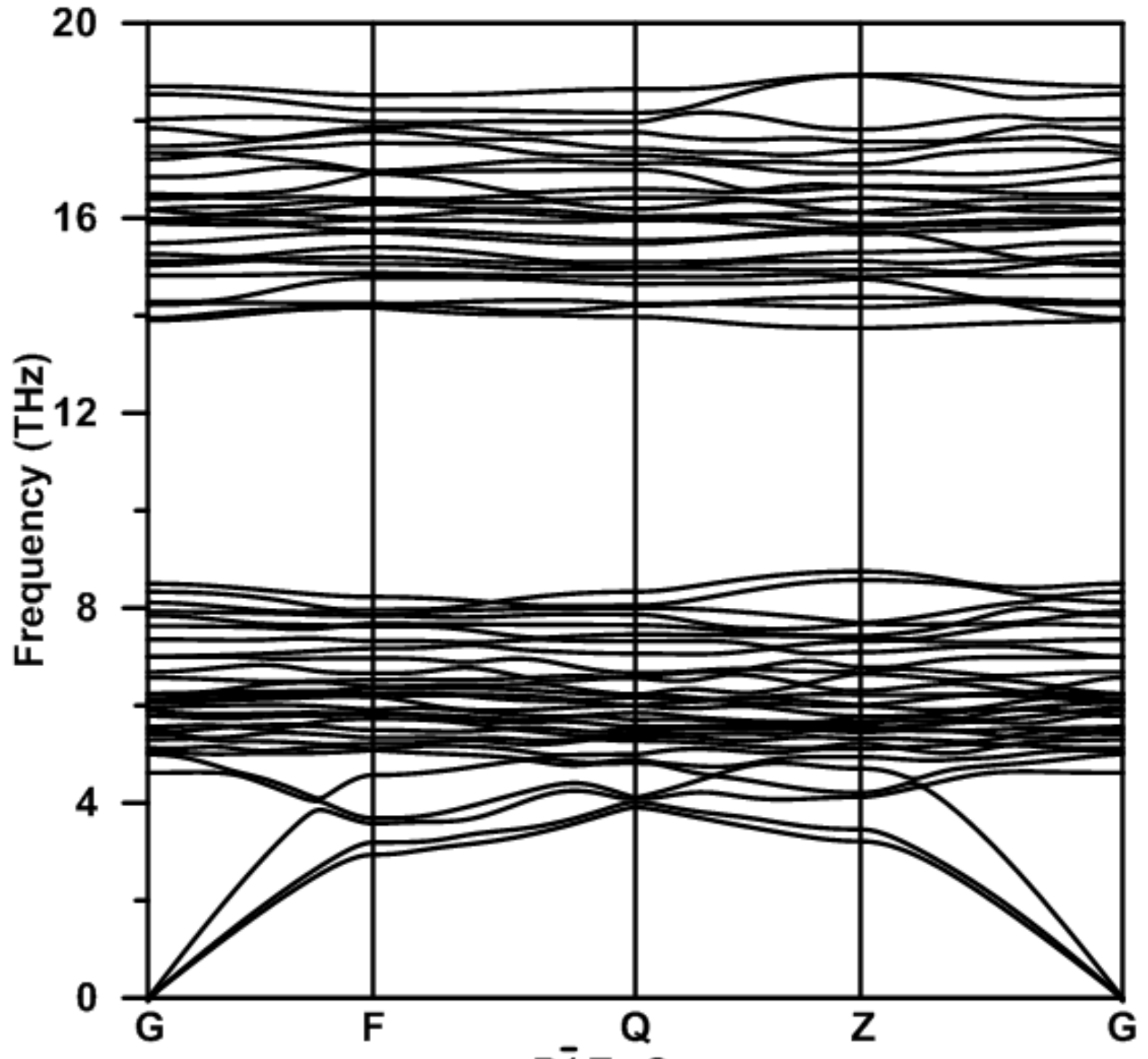

$P\bar{1}$ $Zr_5C_4$

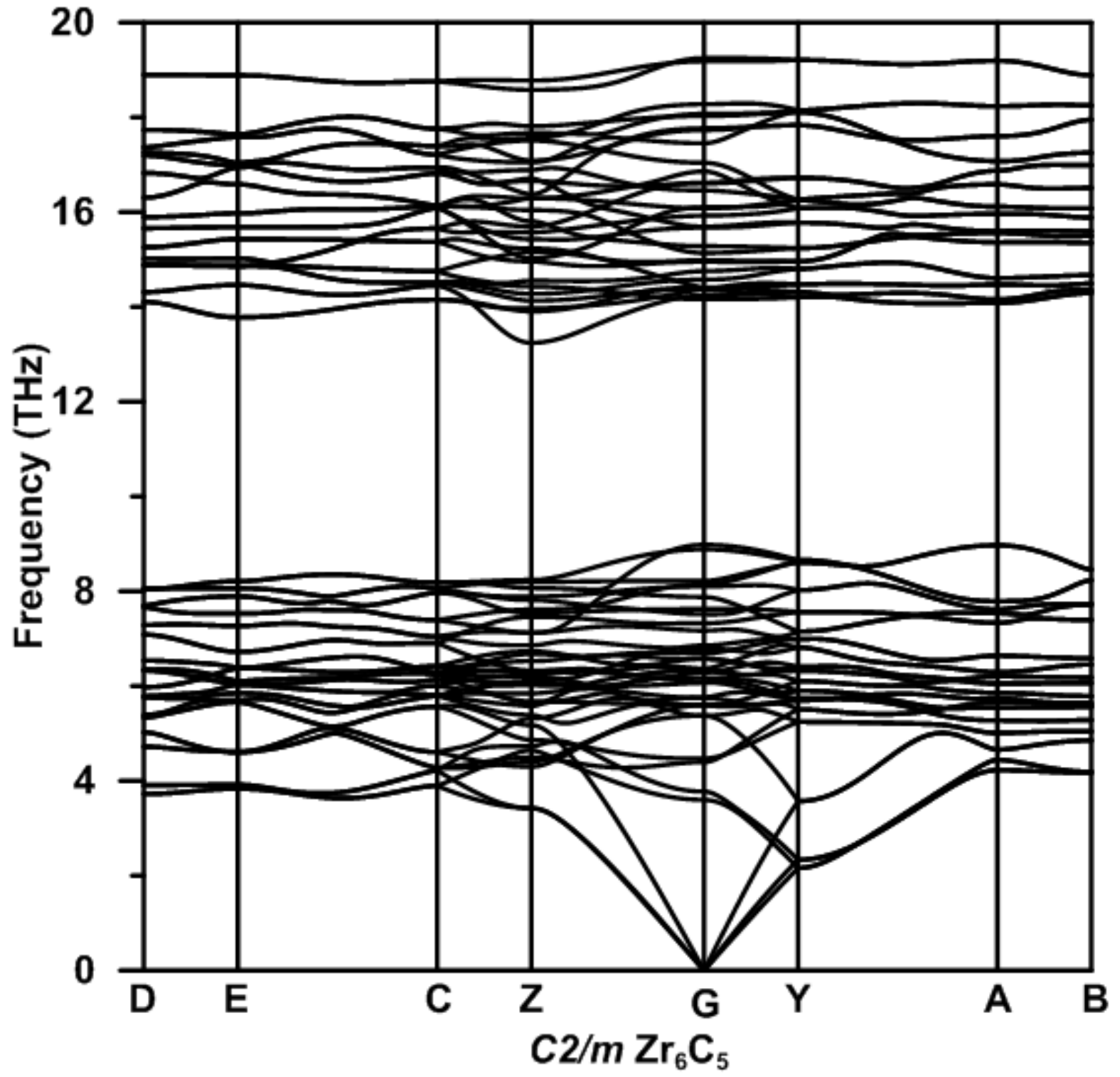

$C2/m$ $Zr_6C_5$

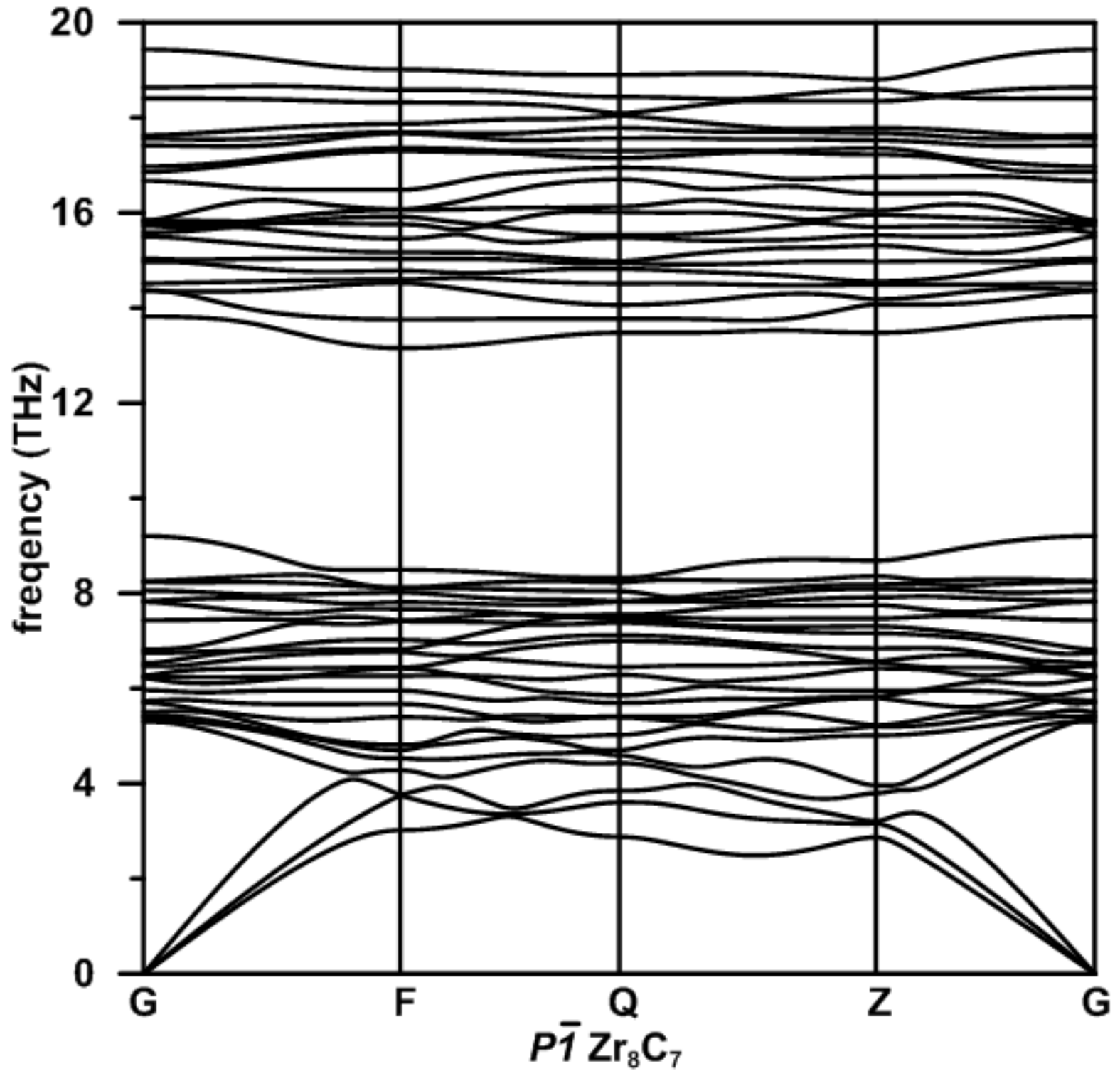

Fig. S3: Phonon dispersion curves computed for all the stable substoichiometric zirconium carbides

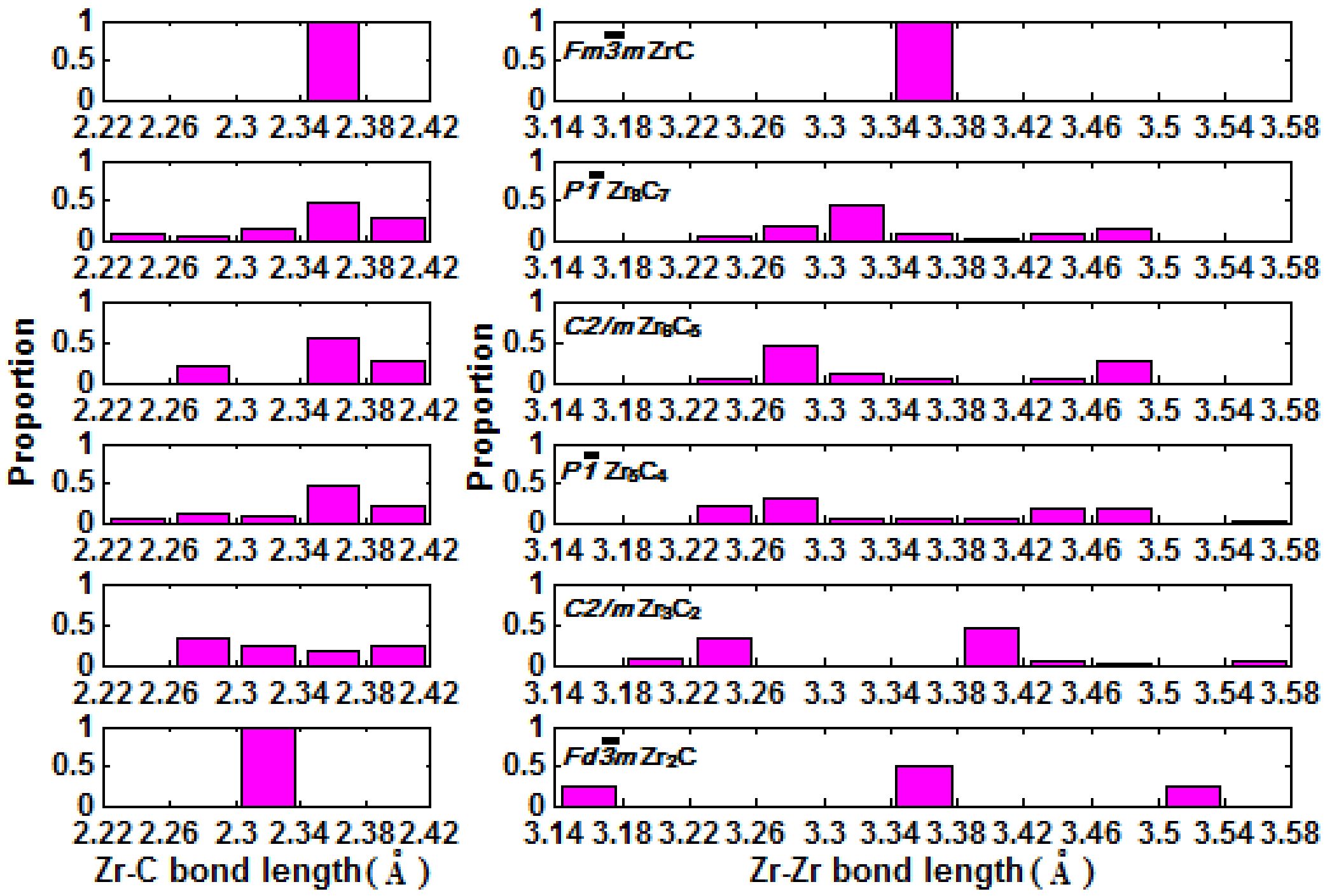

Fig. S4: Distribution of Zr-Zr and Zr-C bonds length (in Å ) in stable zirconium carbides (ZrC, $Zr_8C_7$, $Zr_6C_5$, $Zr_5C_4$, $Zr_3C_2$, and $Zr_2C$).

Table SI. Geometric details of stable zirconium carbides

| ZrC: *Fm-3m*, a = 4.689, V = 103.12 | | | | | | |
|---|---|---|---|---|---|---|
| Element Symbol | No. | Wyckoff positions | x | y | Z | Coordination Number |
| C | 1 | 4*b* | 0.5 | 0.5 | 0.5 | 6 |
| Zr | 1 | 4*a* | 0 | 0 | 0 | 6 |
| $Zr_8C_7$: *P-1*, a = 5.789, b = 5.796, c = 6.701, α = 106.90, β = 89.93, γ = 99.34, V = 212.01 | | | | | | |
| Element Symbol | No. | Wyckoff Positions | x | y | z | Coordination Number |
| C | 1 | 2*i* | 0.2536 | 0.4976 | 0.1251 | 6 |
| C | 2 | 2*i* | 0.2488 | 0.4948 | 0.6235 | 6 |
| C | 3 | 2*i* | 0.0008 | 0.9996 | 0.7476 | 6 |
| C | 4 | 1*d* | 0.5 | 0 | 0 | 6 |
| Zr | 1 | 2*i* | 0.3725 | 0.2624 | 0.3062 | 5 |
| Zr | 2 | 2*i* | 0.8929 | 0.2555 | 0.5653 | 5 |
| Zr | 3 | 2*i* | 0.3706 | 0.2550 | 0.8234 | 5 |
| Zr | 4 | 2*i* | 0.8706 | 0.2454 | 0.0608 | 6 |
| $Zr_6C_5$: *C2/m*, a = 5.754, b = 9.961, c = 6.634, β = 125.17, V = 310.81 | | | | | | |
| Element Symbol | No. | Wyckoff Positions | X | y | z | Coordination Number |
| C | 1 | 4*h* | 0.5 | 0.3340 | 0.5 | 6 |
| C | 2 | 4*g* | 0.5 | 0.3317 | 0 | 6 |
| C | 3 | 2*b* | 0.5 | 0 | 0 | 6 |
| Zr | 1 | 8*i* | 0.4954 | 0.8268 | 0.7539 | 5 |
| Zr | 2 | 4*i* | 0.9792 | 0 | 0.2393 | 5 |
| $Zr_5C_4$: *P-1*, a = 7.487, b = 6.684, c = 5.790, α = 106.78, β = 75.07, γ = 102.84, V = 264.62 | | | | | | |
| Element Symbol | No. | Wyckoff Positions | x | y | z | Coordination Number |
| C | 1 | 2*i* | 0.7994 | 0.3482 | 0.0983 | 6 |
| C | 2 | 2*i* | 0.3995 | 0.5501 | 0.3020 | 6 |
| C | 3 | 2*i* | 0.6038 | 0.9507 | 0.6996 | 6 |
| C | 4 | 2*i* | 0.0015 | 0.7523 | 0.4965 | 6 |
| Zr | 1 | 2*i* | 0.3026 | 0.3559 | 0.5873 | 5 |
| Zr | 2 | 2*i* | 0.3057 | 0.8402 | 0.5906 | 5 |
| Zr | 3 | 2*i* | 0.5127 | 0.2560 | 0.0039 | 5 |
| Zr | 4 | 2*i* | 0.9074 | 0.5424 | 0.7940 | 5 |
| Zr | 5 | 2*i* | 0.1081 | 0.9501 | 0.2140 | 4 |
| $Zr_3C_2$: *C2/m*, a = 5.754, b = 9.943, c = 6.624, β = 125.17, V = 309.80 | | | | | | |
| Element Symbol | No. | Wyckoff Positions | x | y | z | Coordination Number |
| C | 1 | 2*a* | 0 | 0 | 0.5 | 6 |
| C | 2 | 2*a* | 0 | 0 | 0 | 6 |
| C | 3 | 4*g* | 0.5 | 0.1662 | 0 | 6 |
| Zr | 1 | 8*i* | 0.5167 | 0.3391 | 0.7644 | 4 |
| Zr | 2 | 4*i* | 0.5066 | 0 | 0.2439 | 4 |
| $Zr_2C$: *Fd-3m*, a = 9.369, V = 822.31 | | | | | | |
| Element Symbol | No. | Wyckoff Positions | x | y | z | Coordination Number |
| C | 1 | 16*d* | 0.1250 | 0.1250 | 0.6250 | 6 |
| Zr | 1 | 32*e* | 0.3809 | 0.6191 | 0.6191 | 3 |

Table SII The calculated elastic constants $C_{ij}$, bulk modulus $B$ (GPa), shear modulus $G$ (GPa), Pugh's ratio and Vickers hardness $H_v$ (GPa) of Zr-C compounds at ambient conditions.

| Compounds | $Zr_2C$ | $Zr_3C_2$ | $Zr_5C_4$ | $Zr_6C_5$ | $Zr_8C_7$ | ZrC |
|---|---|---|---|---|---|---|
| $C_{11}$ | 200 | 312 | 369 | 419 | 395 | 496 |
| $C_{12}$ | 105 | 88 | 92 | 83 | 110 | 109 |
| $C_{13}$ | | 97 | 101 | 87 | 104 | |
| $C_{14}$ | | | 7 | | 17 | |
| $C_{15}$ | | -9 | 7 | 3 | 4 | |
| $C_{16}$ | | | 9 | | 7 | |
| $C_{22}$ | | 331 | 375 | 386 | 397 | |
| $C_{23}$ | | 86 | 105 | 115 | 106 | |
| $C_{24}$ | | | 6 | | -6 | |
| $C_{25}$ | | -1 | -9 | -1 | 5 | |
| $C_{26}$ | | | -5 | | -7 | |
| $C_{33}$ | | 311 | 353 | 393 | 406 | |
| $C_{34}$ | | | -3 | | -11 | |
| $C_{35}$ | | -3 | -5 | 6 | -14 | |
| $C_{36}$ | | | 1 | | 10 | |
| $C_{44}$ | 93 | 110 | 135 | 127 | 156 | 151 |
| $C_{45}$ | | | 9 | | 4 | |
| $C_{46}$ | | -6 | -9 | -1 | 9 | |
| $C_{55}$ | | 96 | 142 | 162 | 148 | |
| $C_{56}$ | | | -0.2 | | -9 | |
| $C_{66}$ | | 114 | 142 | 131 | 146 | |
| $B$ | 137 | 166 | 188 | 196 | 204 | 238 |
| $G$ | 71 | 109 | 136 | 144 | 148 | 168 |
| $k$ | 0.523 | 0.658 | 0.726 | 0.734 | 0.724 | 0.700 |
| $H_v$ | 8.4 | 16.1 | 21.4 | 22.5 | 22.5 | 23.3 |

Table SIII The calculated formation enthalpy per atom for $ZrC_{1-x}T_x$ (T=B, N, and O; $x$=0, 1/8, 1/6, and 1/5) compounds, defined as $\Delta H(ZrC_{1-x}T_x)=E(ZrC_{1-x}T_x)-E(ZrC_{1-x})- x\,E(T)$.

| Phases | $Zr_8C_7B$ | $Zr_6C_5B$ | $Zr_5C_4B$ | $Zr_8C_7N$ | $Zr_6C_5N$ | $Zr_5C_4N$ | $Zr_8C_7O$ | $Zr_6C_5O$ | $Zr_5C_4O$ |
|---|---|---|---|---|---|---|---|---|---|
| $\Delta H$ | ~0 | ~0 | -0.0111 | -0.1938 | -0.2591 | -0.3135 | -0.3304 | -0.4378 | -0.5226 |